\documentclass[12pt]{article}
\usepackage{times}

\newcounter{lastnote}

\usepackage{color}
\usepackage{longtable}
\usepackage{graphicx}
\usepackage{epsfig}
\usepackage{amssymb}
\usepackage{color}

\newcommand{\beq}{\begin{equation}}
\newcommand{\eeq}{\end{equation}}
\newcommand{\bea}{\begin{eqnarray}}
\newcommand{\eea}{\end{eqnarray}}

\title{Quantum engineering of squeezed states for quantum communication and metrology}

\author{Henning Vahlbruch, Simon Chelkowski, \\ Karsten Danzmann and Roman Schnabel $^{\ast}$\\
\\
\normalsize{Max-Planck-Institut f\"ur Gravitationsphysik
(Albert-Einstein-Institut) and}\\ \normalsize{Institut f\"ur
Gravitationsphysik, Leibniz Universit\"at Hannover}\\
\normalsize{Callinstr. 38, 30167 Hannover, Germany}\\
\\
\normalsize{$^\ast$To whom correspondence should be addressed;
E-mail:  Roman.Schnabel@aei.mpg.de.} }

\date{\today}

\begin{document}

\baselineskip24pt

\maketitle

\textbf{We report the experimental realization of squeezed quantum states of light, tailored for new applications in quantum communication
and metrology. 
Squeezed states in a broad Fourier frequency band down to 1 Hz has been observed for the first time.
Nonclassical properties of light in such a low frequency band is required for high efficiency quantum information storage in electromagnetically induced transparency (EIT) media. The states observed also cover the frequency band of ultra-high precision laser interferometers for gravitational wave detection and can be used to reach the regime of quantum non-demolition interferometry. And furthermore, they cover the frequencies of motions of heavily macroscopic objects and might therefore support the attempts to observe entanglement in our macroscopic world.}


Squeezed states of light constitute a rather peculiar form of light \cite{Wal83}.
Their statistical properties cannot be explained by arrival times of independent photons, and some of their
field measures show less quantum noise than those of the vacuum state, which is the light's ground state with
\emph{zero} average photon number. Squeezed states were first demonstrated by Slusher \emph{et al.} in 1985
\cite{SHYMV85}. Later, experimental techniques for the complete characterization of squeezed states have been
demonstrated \cite{BSM97}. Squeezed states have been used to construct entangled states of light and to demonstrate
quantum teleportation \cite{FSBFKP98, BTBSRBSL03}. Recently, they have been used to engineer Schr\"odinger kitten
states for quantum information networks \cite{OTLG06,NNHMP06}. Applications in the field of metrology and high
precision measurements have been demonstrated in several proof of principle experiments  \cite{MSMBL02,TGBFBL03,VCHFDS05}.
In all these experiments squeezed states on rather short time scales below a microsecond, corresponding to
Fourier frequencies above a megahertz, were employed.
But, many applications require engineering of squeezed states defined on much longer
time scales. High precision laser interferometers can be turned into quantum non-demolition (QND) measurement
devices by employing squeezed states, thereby entering the field of \emph{quantum} metrology. Possible
examples for such interferometers are the currently operated gravitational wave detectors \cite{LIGOGEO04}.
Here stably controlled squeezed states in the gravitational wave detection band of several kilohertz down to
a few hertz will be required.
Squeezed states in the same low frequency regime can be used to create entanglement between light and atoms
to push a \emph{light-atom} interferometer beyond its standard-quantum limit \cite{HOH06}.
In the field of quantum communication, the coherent delay and the storage of nonclassical states of light are
desired. Recently it has been theoretically shown that this can be achieved with electromagnetically induced
transparency (EIT) \cite{AAK04}. Due to the narrow transparency windows, squeezed states at and below
kilohertz frequencies are required \cite{HHGLBBL06}.
In quantum information science, the creation of entangled states of macroscopic objects is of great
scientific interest \cite{Cho03,Rou06}. They allow the study of quantum decoherence and the transformation
from the microscopic quantum world to our macroscopic classical world. It has been theoretically shown that
squeezed states of light can be used to entangle two suspended macroscopic mirrors \cite{PDVABH05}. The
quantum variables of the entanglement will be the positions and momenta of vibrational or pendulum modes of
the mirrors. For massive, macroscopic objects of hundreds of grams or even kilograms, these modes will also
occur at rather low frequencies.
\emph{Quantum engineering} is therefore demanded to prepare new states of light, whose nonclassical
properties reveal themselves on time-scales directly graspable for human beings.

In two pioneering experiments, squeezed states at audio frequencies were first generated \cite{MGBWGML04},
and a coherent control scheme for stable application of such states demonstrated \cite{VCHFDS06}. However, at
Fourier frequencies below one kilohertz the overall noise level of the optical states increased in both
experiments, for reasons that, up to now, had not been understood.
Here we report the observation of squeezed quantum states of light at  Fourier frequencies below the kilohertz regime down
to 1\,Hz. The measured quantum noise levels were in perfect agreement with theoretical predictions. \emph{Parasitic interferences} were identified to be responsible for the previously observed discrepancy at low frequencies.



Quantum states of optical fields can be detected and fully characterized with a \emph{balanced homodyne detector} \cite{BSM97}. The measurement quantities are the field quadratures, e.g. amplitude or phase quadratures, and their variances. A rather impressive property of such a detector is its capability to measure even fluctuations of fields that do not contain \emph{any} photons on average, i.e. fields in so-called vacuum states. These vacuum fluctuations contribute to the zero point energy which is a manifestation of quantum physics after which any oscillator, like a single-mode state of light, cannot have zero energy, otherwise Heisenberg's uncertainty relation would be violated \cite{GerryandKnight04}. A balanced homodyne detector is also a perfect device for the detection of squeezed states of light. In this case the variance of a certain field quadrature is found to be \emph{squeezed} below the variance of the corresponding vacuum field.

\begin{figure}[t]
\centerline{\includegraphics[width=8.6cm]{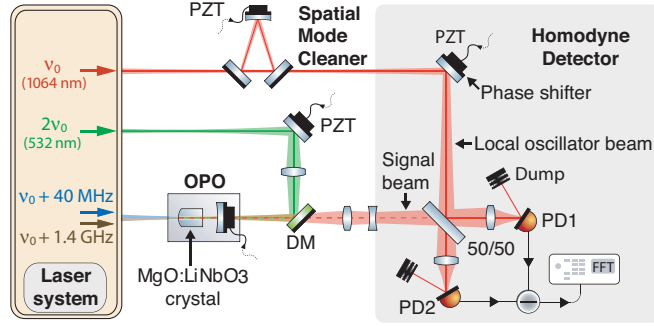}}
  \vspace{0mm}
\caption{Schematic of the experiment utilizing four continuous wave laser beams. Squeezed states at low
Fourier frequencies around carrier frequency $\nu_{0}=c/1064$\,nm (with $c$ the speed of light) were produced
utilizing optical parametric oscillation (OPO). This parametric process was initiated inside a birefringent
crystal made from MgO:LiNbO$_{3}$ by pumping the system with a green laser beam at 532\,nm. Two additional
laser beams were also focussed into the crystal. They served as control beams for piezo-electric length
control of the cavity and phase control of the green beam. The fourth laser beam at 1064\,nm was used as the
local oscillator (LO) to observe the squeezed states by balanced homodyne detection. DM: dichroic mirror; PD:
photo diode; PZT: piezo-electric transducer for positioning of mirrors. }
  \label{experiment}
\end{figure}

In balanced homodyne detection dim quantum states of the signal beam are interfered with an intense auxiliary
laser beam (local oscillator, LO) on a beam splitter with 50\% power reflectivity,
as shown in Fig.\,\ref{experiment}. The interference with the LO leads to an optical amplification of the measured signal field quadrature by a large factor, which is necessary to reach levels far above electronic noise of photodiodes and subsequent electronic circuits. Each output field from
the 50/50 beam splitter is focussed onto a semiconductor photodiode.
The final signal is derived from the difference of the two photocurrents, which is
spectrally analyzed, for example using a Fast-Fourier-Transformation (FFT). Please note that noise
contributions from the LO beam cancel in balanced homodyne detection.
If the signal field and the LO interfere \emph{in} phase, the balanced homodyne detector measures the signal
state's \emph{amplitude} quadrature $\hat q_1 (\Omega, \Delta\Omega, t)$ where $\Omega/2\pi$ is the Fourier
frequency and $\Delta\Omega/2\pi$ is the detection resolution bandwidth (RBW).
If the LO's optical path length is changed by a quarter wave length 
the homodyne detector measures the phase quadrature $\hat q_2
(\Omega, \Delta\Omega, t)$. The amplitude, together with the phase quadrature, form a set of two
non-commuting observables. The simultaneous precise knowledge of both their values is limited by Heisenberg's
uncertainty relation.

\begin{figure}[t]
\centerline{
  \includegraphics[width=5.6cm]{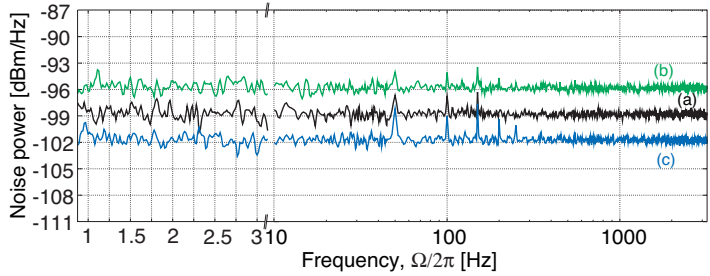}}
  \vspace{0mm}
  \caption{Noise powers (variances) of field quadratures  for a spectrum of vacuum states at different sideband frequencies.
    Three different local oscillator
  powers were used:  (a) 464\,$\mu$W, (b) 928\,$\mu$W, and (c) 232\,$\mu$W.  All traces are pieced together from five FFT frequency windows: 0.8--3.2\,Hz, 10--50\,Hz, 50--200\,Hz, 200--800\,Hz, 800\,Hz--3.2\,kHz with resolution bandwidths (RBW) of $\Delta\Omega/2\pi$=15.63\,mHz, 250\,mHz, 1\,Hz, 2\,Hz and 4\,Hz, respectively.
  Each measurement point is the averaged root mean square value of 75, 100, 100, 400 and 400 measurements,
  again respectively for the five FFT windows.
    Peaks at 50\,Hz and harmonics were due to the electric mains.
    Here, the electronic dark noise has been subtracted from the data.}
  \label{shotnoise}
\end{figure}

Since the vacuum state sets the reference for squeezed states, we first present measurements of vacuum
fluctuations. Vacuum states were produced by carefully blocking the signal input beam of the balanced homodyne
detector without introducing scattered fields, while keeping the LO beam. Fig.\,\ref{shotnoise} shows the spectral noise powers of optically amplified
vacuum fluctuations of an optical field mode at 1064\,nm  for three different LO powers measured with our
homodyne detector. We plot the spectral decomposition of the vacuum noise for sideband frequencies $\Omega /
2 \pi$ between 800\,mHz and 3.2\,kHz.
For all three LO powers the signal port of our homodyne detector was empty
and the observation of pure vacuum noise was confirmed in the following way.
Firstly, the measured spectral noise powers scaled linearly with LO power. Secondly, the noise spectra were white, i.e.\,independent of Fourier frequency $\Omega / 2 \pi$. Both properties of vacuum noise are predicted from theory and were clearly found in our
observations as presented in Fig.\,\ref{shotnoise}. Thirdly, independent measurements of the amplification
factors of all electronic components used confirmed \emph{quantitatively} within $\pm$0.5\,dB uncertainty,
that indeed pure vacuum noise was observed. We point out that the measurement of the frequency interval in Fig.\,\ref{shotnoise} lasted for more than half an hour for each LO power, thereby
demonstrating long term stability of our detector-setup.
Our results in Fig.\,\ref{shotnoise} represent the first successful detection of vacuum states at
sub-audio and audio frequencies below one kHz with a white (flat) spectrum.



We now discuss how an optical field with less fluctuations than vacuum states at
sideband frequencies down to 1\,Hz was generated. A simplified schematic of the experiment is shown in
Fig.\,\ref{experiment}. In total four linearly polarized, continuous-wave laser beams were used. They
originated from two monolithic non-planar Nd:YAG continuous wave ring laser devices operating at 1064\,nm.
One of the laser devices provided the reference frequency $\nu_0$. Parts of this field were converted into
laser radiation at 2$\nu_0$ and $\nu_0$+40\,MHz, utilizing a birefringent crystal and an acousto-optical
modulator, respectively, (crystal and modulator are not shown in Fig.\,\ref{experiment}). The second laser
device was phase locked to the first one and was operated at a frequency of $\nu_0$+1.4\,GHz.
The frequency-doubled and frequency-shifted beams were all mode-matched into a nonlinear standing-wave cavity
which housed a 6.5\,mm long, 7\% doped MgO:LiNbO$_{3}$ crystal.
The frequency-doubled beam was horizontally polarized and served as the pump field for type I degenerate
optical parametric oscillation (OPO) below threshold.
This process deamplified the quadratures, $\hat q_1$, of all sideband fields of the vertically polarized cavity resonance
mode at the fundamental frequency $\nu_0$, defined at Fourier frequencies up to the cavity linewidth of
approximately 27\,MHz. Due to the OPO process and in accordance with Heisenberg's uncertainty relation, the
orthogonal quadratures $\hat q_2$ were then complementarily amplified.

Since no laser radiation around the frequency $\nu_0$ was injected into the OPO cavity, the parametric
process only acted on vacuum fluctuations and produced squeezed vacuum states for all frequencies within the
OPO cavity linewidth. The squeezed states were coupled out in the backwards direction of the injected pump
field via a dichroic mirror and were sent to the balanced homodyne detector.
The frequency-shifted beams were used to sense the two essential degrees of freedom of our experiment. The
first one was the length of the OPO cavity. Only if the OPO cavity was kept on resonance for the reference
frequency $\nu_0$ (half the pump field frequency) squeezed states were produced. The second one was the
relative phase of the second harmonic OPO pump field with respect to the local oscillator of the downstream
experiment. Only when this relative phase was zero could the squeezed amplitude quadrature $\hat q_1$ be
observed by our homodyne detector. The OPO cavity length was sensed by the beam at $\nu_0$+1.4\,GHz. This
beam was orthogonally polarized with respect to the squeezed cavity mode and its frequency shift exactly
compensated the birefringence of the OPO crystal. The phase of the pump field was sensed by the beam at
$\nu_0$+40\,MHz; the procedure has been described in much detail in \cite{VCHFDS06,CVDS07}. The information from the
two frequency shifted beams were then used to fix both degrees of freedom by precisely positioning of the
relevant mirrors using piezo-electric transducers and electronic feedback loops. The special and important
property of the scheme was that solely \emph{frequency shifted} beams were used. If a field directly at the
reference frequency $\nu_0$ was used, an observation of squeezed states at low frequencies would not be
possible, as shown in \cite{MGBWGML04}.

\begin{figure}
  \centerline{\includegraphics[width=5.6cm]{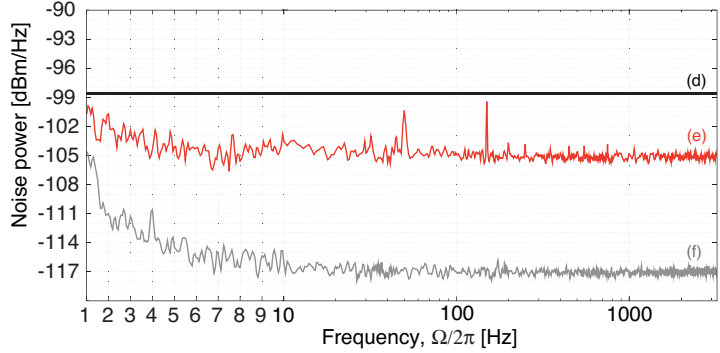}}
  \vspace{0mm}
  \caption{Trace (d) shows the (theoretical) vacuum noise level of our homodyne detector with 464\,$\mu$W local oscillator power as confirmed in Fig.\,\ref{shotnoise}. Trace (e) shows the noise powers of squeezed states measured with the same local oscillator power. Squeezed vacuum noise
  was observed throughout the complete spectrum from 1\,Hz to above 3\,kHz.
  A nonclassical noise suppression of up to 6.5\,dB below vacuum noise was observed here.
  FFT frequency windows: 1--10\,Hz, 10--50\,Hz, 50--200\,Hz, 200--800\,Hz, 800\,Hz--3.2\,kHz.
  RBW: 62.5\,mHz, 250\,mHz, 1\,Hz, 2\,Hz and 4\,Hz. Averages: 30, 100, 100, 400 and 400.
  The dark noise (f) was not subtracted from the measured data.}
  \label{Squeezing}
\end{figure}

The observed variances of squeezed states between 1\,Hz and 3.2\,kHz are presented in Fig.\,\ref{Squeezing},
trace (e). For a significant portion of the shown spectrum the quantum noise variance was squeezed by an
average value of 6.5\,dB, i.e. 4.5 times smaller than the vacuum noise variance, trace (d). This value
exactly matched the theoretical prediction for our experiment as discussed below. Trace (f) shows the
dark-noise contribution of the homodyne detector itself.

The strength of the observed squeezing was limited by the finite optical parametric gain of the OPO cavity and the total
optical loss the squeezed states suffered before being detected. For the measurements shown in
Fig.\,\ref{Squeezing} a gain of $g\!=$12$\pm$0.5 was used and which was achieved with a pump power of
100\,mW. Optical losses inside the OPO, in the beam path and in the homodyne detector added up to
$\,l\!=$0.15$\pm$0.04. The error bar was mainly due to the uncertainty in the quantum efficiency of the
homodyne detector photodiodes (PD1, PD2 in Fig.\,\ref{experiment}, type: ETX500). Both values ($g$, $l$) were
deduced from separate measurements. They can be used to estimate the expected strength of squeezing (for
Fourier frequencies much smaller than the OPO cavity line width) and indeed provide the observed value:
$-10\cdot$log$_{10}(l+(1-l)/g) \!\approx$6.5\,dB.

Compared with other experiments at audio frequencies, see for example Ref.\,\cite{VCHFDS06}, the squeezing
strength reported here has been considerably increased. The improvement was enabled by a spatial mode
cleaning cavity in the LO beam path right before the homodyne detector 50/50 beam splitter. With this cavity
the spatial overlap, i.e. the interference contrast between LO and signal beam, could be increased up to a
fringe visibility of 98.3\%.
We note that the squeezed variances shown in Fig.\,\ref{Squeezing} represent typical results with high long
term stability of the setup. With an increased second harmonic pump power the classical OPO gain changed from
approximately 12 to 40$\pm$4 and up to 7.2\,dB squeezing could be directly observed (7.5\,dB when the dark
noise was subtracted). In this regime stable operation of the OPO cavity was still possible for several
minutes. However, for longer periods thermal fluctuations inside the OPO crystal due to absorption and power
fluctuations of the pump beam moved the frequency shift between the fundamental mode and the coherent control
beam away from its required value of 1.4\,GHz. Long term stable operation with high parametric gain should be
possible with an electro-optical stabilization of the second harmonic pump power.
At sub-audio frequencies below 5\,Hz, trace (e) shows degraded squeezing strengths. An
averaged value of 1.5\,dB below vacuum noise was observed at 1\,Hz. The degradation
was partly due to a rising electronic dark noise level (trace (f)). By subtracting the dark noise, the
nonclassical noise suppression recovered to 3.5$\pm$0.5\,dB which was, however, still significantly lower
than 6.5\,dB as observed at other frequencies. This degradation was due to remnant parasitic interferences as
described in the next paragraph.



The generation and observation of squeezed vacuum states of light at (sub-) audio frequencies reported here,
were made possible by a significant reduction of \emph{parasitic interferences}. Parasitic interferences are
typically produced by moving surfaces that scatter photons into the low frequency detection band of the
homodyne detector.
Scattered light fields were identified to originate from the micro-roughness of the optical surfaces,
non-perfect anti-reflection coatings and residual transmissions of high reflection mirrors. They were partly
transmitted through the OPO cavity and entered our homodyne detector via its signal port. Other scattered
fields were found to directly hit the photodiodes of the homodyne detector from other directions. In all
cases these fields sensed multiple scattering processes from various optics and optic mounts. The scattering
surfaces of these objects continuously moved through thermal expansion and acoustic vibrations. During the
scattering processes optical sideband fields were produced at corresponding Fourier frequencies and higher
harmonics.
For any such scattered field which interfered with the homodyne local oscillator the whole setup acted like a
sensitive interferometer measuring the motions of the scattering surfaces. These parasitic interferences
easily masked the vacuum noise. Given a great number of such sources of scattered light with increasing
motion amplitudes towards lower frequencies but low mechanical quality factors, a rather smooth monotonic
increase was produced as observed for example in \cite{VCHFDS06}.
We reduced the parasitic interferences firstly, by carefully shielding our homodyne detector against
scattered light fields and secondly, by reducing air turbulence, vibrations and temperature fluctuations of
our experiment.
The success of our experimental improvements proved, that the previous limitations to the observation of
squeezed states were purely optical and were not due to any type of noise related to the photodiodes'
semiconductor material. We could further validate that noise of the LO, amplified by a residual signal port
field at optical frequency $\nu_0$, gave no significant contribution to the experiment reported in
\cite{VCHFDS06}. Therefore, previously reported homodyne detector noise spectra were, most likely, not
vacuum noise dominated below one kilohertz. In the same way squeezed variance spectra at these frequencies
were also contaminated by additional noise from scattered fields.



In conclusion, our experiments expanded the range of Fourier frequencies at which squeezed states of light have been
engineered by two decades. We reached the frequency regime in which vibrations and rotations of real
macroscopic objects of human's everyday experience occur opening new avenues for
high precision optical measurements of quantum metrology, as well as for quantum memories as a key element of
many quantum communication protocols.
A nonclassical noise suppression of up to 7.2\,dB below vacuum noise variance at audio Fourier frequencies was achieved.
Based upon our investigations we believe that squeezed states at even
lower frequencies with also higher degrees of squeezing can be allocated by further protection against
parasitic interferences and reduction of optical losses.

We thank B.\,Hage and A.\,Franzen for valuable contributions to the experiment. This work has been supported
by the Deutsche Forschungsgemeinschaft and is part of Sonderforschungsbereich 407.

\end{document}